\documentclass[aip,amsmath,amssymb,twocolumn,]{revtex4}                

\usepackage{textcase}
\usepackage{graphicx}
\usepackage{dcolumn}
\usepackage{bm}
\usepackage{amsmath,amsthm,amssymb,mathrsfs}

\begin{document}


\title{Large amplitude oscillation of magnetization in spin-torque oscillator stabilized by field-like torque}

\author{Tomohiro Taniguchi${}^{1}$, Sumito Tsunegi${}^{2}$, Hitoshi Kubota${}^{1}$, and Hiroshi Imamura${}^{1}$}

\affiliation{ 
${}^{1}$National Institute of Advanced Industrial Science and Technology (AIST), Spintronics Research Center, Tsukuba 305-8568, Japan,
${}^{2}$Unit\'e Mixte de Physique CNRS/Thales and Universit\'e Paris Sud 11, 1 av. A. Fresnel, Palaiseau, France. 
}

\date{\today}%


\begin{abstract}
  Oscillation frequency of spin torque oscillator with a perpendicularly magnetized free layer and an in-plane magnetized pinned layer 
  is theoretically investigated by taking into account the field-like torque. 
  It is shown that the field-like torque plays an important role in finding the balance between 
  the energy supplied by the spin torque and the dissipation due to the damping, 
  which results in a steady precession. 
  The validity of the developed theory is confirmed 
  by performing numerical simulations 
  based on the Landau-Lifshitz-Gilbert equation. 
\end{abstract}

\maketitle


Spin torque oscillator (STO) has attracted much attention 
as a future nanocommunication device 
because it can produce a large emission power ($>1$ $\mu$W), 
a high quality factor ($>10{}^{3}$), 
a high oscillation frequency ($>1$ GHz), 
a wide frequency tunability ($>3$ GHz), 
and a narrow linewidth ($<10^{2}$ kHz) 
\cite{kiselev03,rippard04,houssameddine07,bonetti09,zeng13,kubota13,maehara13,tsunegi14,dussaux14}. 
In particular, STO with a perpendicularly magnetized free layer 
and an in-plane magnetized pinned layer has been developed 
after the discovery of an enhancement of perpendicular anisotropy of CoFeB free layer 
by attaching MgO capping layer \cite{yakata09,ikeda10,kubota12}. 
In the following, we focus on this type of STO. 
We have investigated the oscillation properties of this STO 
both experimentally \cite{kubota13,tsunegi14a} and theoretically \cite{taniguchi13,taniguchi14a}. 
An important conclusion derived in these studies was that 
field-like torque is necessary to excite the self-oscillation 
in the absence of an external field, 
nevertheless the field-like torque is typically 
one to two orders of magnitude smaller than the spin torque \cite{slonczewski96,berger96,tulapurkar05}. 
We showed this conclusion by performing 
numerical simulations based on the Landau-Lifshitz-Gilbert (LLG) equation \cite{taniguchi14a}.

This paper theoretically proves the reason why the field-like torque is necessary 
to excite the oscillation 
by using the energy balance equation 
\cite{apalkov05,bertotti06,dykman12,newhall13,pinna13,pinna14,taniguchi13a,taniguchi13b,taniguchi14b}. 
An effective energy including the effect of the field-like torque is introduced. 
It is shown that introducing field-like torque is crucial in 
finding the energy balance between the spin torque and the damping, 
and as a result to stabilize a steady precession. 
A good agreement with the LLG simulation 
on the current dependence of the oscillation frequency 
shows the validity of the presented theory. 


\begin{figure}
\centerline{\includegraphics[width=0.6\columnwidth]{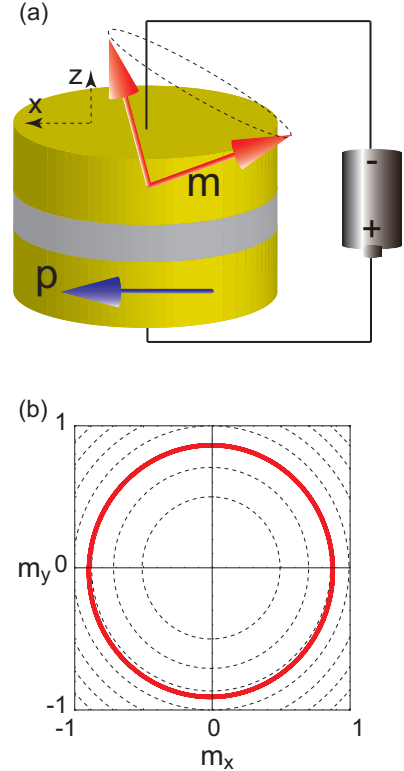}}\vspace{-3.0ex}
\caption{
         (a) Schematic view of the system. 
         (b) Schematic views of the contour plot of the effective energy map (dotted), Eq. (\ref{eq:effective_energy}), 
             and precession trajectory in a steady state with $I=1.6$ mA (solid). 
         \vspace{-3ex}}
\label{fig:fig1}
\end{figure}



The system under consideration is schematically shown in Fig. \ref{fig:fig1} (a). 
The unit vectors pointing in the magnetization directions of the free and pinned layers are 
denoted as $\mathbf{m}$ and $\mathbf{p}$, respectively. 
The $z$-axis is normal to the film-plane, 
whereas the $x$-axis is parallel to the pinned layer magnetization. 
The current $I$ is positive when electrons flow from the free layer to the pinned layer. 
The LLG equation of the free layer magnetization $\mathbf{m}$ is 
\begin{equation}
\begin{split}
  \frac{d \mathbf{m}}{dt}
  =&
  -\gamma
  \mathbf{m}
  \times
  \mathbf{H}
  -
  \gamma
  H_{\rm s}
  \mathbf{m}
  \times
  \left(
    \mathbf{p}
    \times
    \mathbf{m}
  \right)
\\
  &-
  \gamma 
  \beta 
  H_{\rm s}
  \mathbf{m}
  \times
  \mathbf{p}
  +
  \alpha 
  \mathbf{m}
  \times
  \frac{d \mathbf{m}}{dt}, 
  \label{eq:LLG}
\end{split}
\end{equation}
where $\gamma$ is the gyromagnetic ratio. 
Since the external field is assumed to be zero throughout this paper, 
the magnetic field $\mathbf{H}=(H_{\rm K}-4\pi M)m_{z}\mathbf{e}_{z}$ consists of the perpendicular anisotropy field only, 
where $H_{\rm K}$ and $4\pi M$ are the crystalline and shape anisotropy fields, respectively. 
Since we are interested in the perpendicularly magnetized free layer, $H_{\rm K}$ should be larger than $4\pi M$. 
The second and third terms on the right-hand-side of Eq. (\ref{eq:LLG}) are 
the spin torque and field-like torque, respectively. 
The spin torque strength, $H_{\rm s}=\hbar \eta I/[2e (1 + \lambda \mathbf{m}\cdot\mathbf{p}) MV]$, 
includes the saturation magnetization $M$ and volume $V$ of the free layer. 
The spin polarization of the current and the dependence of the spin torque strength on the relative angle of the magnetizations 
are characterized in respective by $\eta$ and $\lambda$ \cite{taniguchi13}. 
According to Ref. \cite{taniguchi14a}, $\beta$ should be negative 
to stabilize the self-oscillation. 
The values of the parameters used in the following calculations are 
$M=1448$ emu/c.c., 
$H_{\rm K}=20.0$ kOe, 
$V=\pi \times 60 \times 60 \times 2$ nm${}^{3}$, 
$\eta=0.54$, 
$\lambda=\eta^{2}$, 
$\beta=-0.2$, 
$\gamma=1.732 \times 10^{7}$ rad/(Oe$\cdot$s), 
and $\alpha=0.005$, respectively \cite{kubota13,taniguchi14a}. 
The critical current of the magnetization dynamics for $\beta=0$ is 
$I_{\rm c}=[4 \alpha eMV/(\hbar \eta \lambda)](H_{\rm K}-4\pi M) \simeq 1.2$ mA, 
where Ref. \cite{taniguchi14a} shows that the effect of $\beta$ on the critical current is negligible. 
When the current magnitude is below the critical current, 
the magnetization is stabilized at $m_{z}=1$. 


In the oscillation state, 
the energy supplied by the spin torque balances the dissipation due to the damping. 
Usually, the energy is the magnetic energy density defined as 
$E=-M \int d \mathbf{m} \cdot \mathbf{H}$ \cite{lifshitz80}, 
which includes the perpendicular anisotropy energy only, $-M(H_{\rm K}-4\pi M) m_{z}^{2}/2$, 
in the present model. 
The first term on the right-hand-side of Eq. (\ref{eq:LLG}) can be expressed as 
$-\gamma \mathbf{m} \times [-\partial E/\partial (M \mathbf{m})]$. 
However, Eq. (\ref{eq:LLG}) indicates that 
an effective energy density,
\begin{equation}
  E_{\rm eff}
  =
  -\frac{M (H_{\rm K}-4\pi M)}{2}
  m_{z}^{2}
  -
  \frac{\beta \hbar \eta I}{2e \lambda V}
  \log 
  \left(
    1
    +
    \lambda 
    \mathbf{m}
    \cdot
    \mathbf{p}
  \right), 
  \label{eq:effective_energy}
\end{equation}
should be introduced because 
the first and third terms on the right-hand-side of Eq. (\ref{eq:LLG}) 
can be summarized as $-\gamma \mathbf{m}\times [-\partial E_{\rm eff}/\partial (M \mathbf{m})]$. 
Here, we introduce an effective magnetic field 
$\bm{\mathcal{H}}=-\partial E_{\rm eff}/\partial (M \mathbf{m})
  =(\beta \hbar \eta I/[2e (1 + \lambda m_{x}) MV],0,(H_{\rm K}-4\pi M)m_{z})$.
Dotted line in Fig. \ref{fig:fig1} (b) schematically shows 
the contour plot of the effective energy density $E_{\rm eff}$ projected to the $xy$-plane, 
where the constant energy curves slightly shift along the $x$-axis 
because the second term in Eq. (\ref{eq:effective_energy}) breaks the axial symmetry of $E$. 
Solid line in Fig. \ref{fig:fig1} (b) shows the precession trajectory of the magnetization 
in a steady state with $I=1.6$ mA 
obtained from the LLG equation. 
As shown, the magnetization steadily precesses practically on a constant energy curve of $E_{\rm eff}$. 
Under a given current $I$, 
the effective energy density $E_{\rm eff}$ 
determining the constant energy curve of the stable precession is obtained 
by the energy balance equation \cite{taniguchi14b} 
\begin{equation}
  \alpha 
  \mathscr{M}_{\alpha}(E_{\rm eff})
  -
  \mathscr{M}_{\rm s}(E_{\rm eff})
  =
  0. 
  \label{eq:balance_equation}
\end{equation}
In this equation, $\mathscr{M}_{\alpha}$ and $\mathscr{M}_{\rm s}$, 
which are proportional to the dissipation due to the damping 
and energy supplied by the spin torque during a precession on the constant energy curve, are defined as 
\cite{taniguchi13,taniguchi13a,taniguchi13b,taniguchi14b} 
\begin{equation}
  \mathscr{M}_{\alpha}
  =
  \gamma^{2}
  \oint
  dt 
  \left[
    \bm{\mathcal{H}}^{2}
    -
    \left(
      \mathbf{m}
      \cdot
      \bm{\mathcal{H}}
    \right)^{2}
  \right],
  \label{eq:Melnikov_alpha}
\end{equation}
\begin{equation}
  \mathscr{M}_{\rm s}
  =
  \gamma^{2}
  \oint
  dt 
  H_{\rm s}
  \left[
    \mathbf{p}
    \cdot
    \bm{\mathcal{H}}
    -
    \left(
      \mathbf{m}
      \cdot
      \mathbf{p}
    \right)
    \left(
      \mathbf{m}
      \cdot
      \bm{\mathcal{H}}
    \right)
    -
    \alpha
    \mathbf{p}
    \cdot
    \left(
      \mathbf{m}
      \times
      \bm{\mathcal{H}}
    \right)
  \right]. 
  \label{eq:Melnikov_spin}
\end{equation}
The oscillation frequency on the constant energy curve 
determined by Eq. (\ref{eq:balance_equation}) is given by 
\begin{equation}
  f
  =
  1 \big/
  \oint dt. 
  \label{eq:frequency}
\end{equation}
Since we are interested in zero-field oscillation, 
and from the fact that the cross section of STO in experiment \cite{kubota13} is circle, 
we neglect external field $\mathbf{H}_{\rm ext}$ 
or with in-plane anisotropy field $H_{\rm K}^{\rm in-plane}m_{x}\mathbf{e}_{x}$. 
However, the above formula can be expanded to system with such effects 
by adding these fields to $\bm{\mathcal{H}}$ 
and terms $-M\mathbf{H}_{\rm ext}\cdot\mathbf{m}-MH_{\rm K}^{\rm in-plane}m_{x}^{2}/2$ to the effective energy.



\begin{figure}
\centerline{\includegraphics[width=0.7\columnwidth]{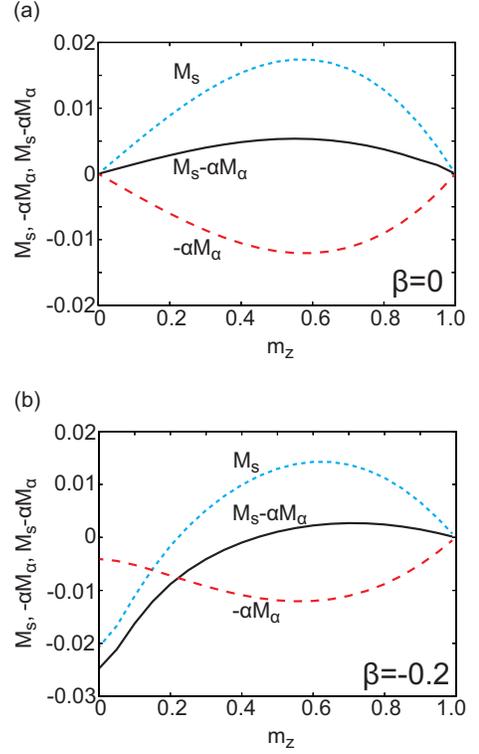}}\vspace{-3.0ex}
\caption{
         Dependences of $\mathscr{M}_{\rm s}$, $-\alpha\mathscr{M}_{\alpha}$, and their difference $\mathscr{M}_{\rm s}-\mathscr{M}_{\alpha}$ 
         normalized by $\gamma (H_{\rm K}-4\pi M)$ 
         on $m_{z}$ ($0 \le m_{z} < 1$)
         for (a) $\beta=0$, and (b) $\beta=-0.2$, 
         where $I=1.6$ mA. 
         \vspace{-3ex}}
\label{fig:fig2}
\end{figure}



In the absence of the field-like torque ($\beta=0$), 
i.e., $E_{\rm eff}=E$, 
there is one-to-one correspondence between the energy density $E$ and $m_{z}$. 
Because an experimentally measurable quantity is the magnetoresistance 
proportional to $(R_{\rm AP}-R_{\rm P}){\rm max}[\mathbf{m}\cdot\mathbf{p}] \propto {\rm max}[m_{x}] = \sqrt{1-m_{z}^{2}}$, 
it is suitable to calculate Eq. (\ref{eq:balance_equation}) as a function of $m_{z}$, instead of $E$, 
where $R_{\rm P(AP)}$ is the resistance of STO in the (anti)parallel alignment of the magnetizations. 
Figure \ref{fig:fig2} (a) shows
dependences of $\mathscr{M}_{\rm s}$, $-\alpha\mathscr{M}_{\alpha}$, and their difference $\mathscr{M}_{\rm s}-\alpha\mathscr{M}_{\alpha}$ 
on $m_{z}$ ($0 \le m_{z} < 1$)
for $\beta=0$, 
where $\mathscr{M}_{\rm s}$ and $\mathscr{M}_{\alpha}$ are normalized by $\gamma (H_{\rm K}-4\pi M)$. 
The current is set as $I=1.6$ mA ($>I_{\rm c}$). 
We also show $\mathscr{M}_{\rm s}$, $-\alpha\mathscr{M}_{\alpha}$, and their difference $\mathscr{M}_{\rm s}-\alpha\mathscr{M}_{\alpha}$ for $\beta=-0.2$ 
in Fig. \ref{fig:fig2} (b), 
where $m_{x}$ is set as $m_{x}=-\sqrt{1-m_{z}^{2}}$. 
Because $-\alpha \mathscr{M}_{\alpha}$ is proportional to the dissipation due to the damping, 
$-\alpha \mathscr{M}_{\alpha}$ is always $-\alpha \mathscr{M}_{\alpha} \le 0$. 
The implications of Figs. \ref{fig:fig2} (a) and (b) are as follows. 
In Fig. \ref{fig:fig2} (a), $\mathscr{M}_{\rm s}-\alpha\mathscr{M}_{\alpha}$ is always positive. 
This means that the energy supplied by the spin torque is always larger than 
the dissipation due to the damping, 
and thus, the net energy absorbed in the free layer is positive. 
Then, starting from the initial equilibrium state ($m_{z}=1$), 
the free layer magnetization moves to the in-plane $m_{z}=0$, 
as shown in Ref. \cite{taniguchi13}. 
On the other hand, in Fig. \ref{fig:fig2} (b), 
$\mathscr{M}_{\rm s}-\alpha\mathscr{M}_{\alpha}$ is positive from $m_{z}=1$ to a certain $m_{z}^{\prime}$, 
whereas it is negative from $m_{z}^{\prime}$ to $m_{z}=0$ 
($m_{z}^{\prime} \simeq 0.4$ in the case of Fig. \ref{fig:fig2} (b)). 
This means that, 
starting from $m_{z}=1$, the magnetization can move to a point $m_{z}^{\prime}$ 
because the net energy absorbed by the free layer is positive, 
which drives the magnetization dynamics. 
However, the magnetization cannot move to the film plane ($m_{z}=0$) 
because the dissipation overcomes the energy supplied by the spin torque 
from $m_{z}=m_{z}^{\prime}$ to $m_{z}=0$. 
Then, a stable and large amplitude precession is realized on a constant energy curve. 



\begin{figure}
\centerline{\includegraphics[width=0.7\columnwidth]{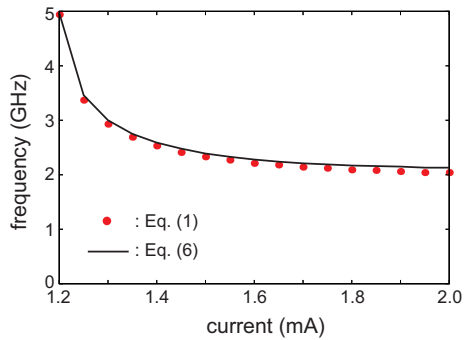}}\vspace{-3.0ex}
\caption{
         Current dependences of peak frequency of $|m_{x}(f)|$ obtained from Eq. (\ref{eq:LLG}) (red circle), 
         and the oscillation frequency estimated by using (\ref{eq:frequency}) (solid line). 
         \vspace{-3ex}}
\label{fig:fig3}
\end{figure}

We confirm the accuracy of the above formula 
by comparing the oscillation frequency estimated by Eq. (\ref{eq:frequency}) 
with the numerical solution of the LLG equation, Eq. (\ref{eq:LLG}). 
In Fig. \ref{fig:fig3}, we summarize the peak frequency of $|m_{x}(f)|$ for $I=1.2-2.0$ mA (solid line), 
where $m_{x}(f)$ is the Fourier transformation of $m_{x}(t)$. 
We also show the oscillation frequency estimated from Eq. (\ref{eq:frequency}) by the dots. 
A quantitatively good agreement is obtained, 
guaranteeing the validity of Eq. (\ref{eq:frequency}). 



In conclusion, 
we developed a theoretical formula to 
evaluate the zero-field oscillation frequency of STO 
in the presence of the field-like torque. 
Our approach was based on the energy balance equation 
between the energy supplied by the spin torque 
and the dissipation due to the damping. 
An effective energy density was introduced to take into account the effect of the field-like torque. 
We discussed that introducing field-like torque is necessary to 
find the energy balance between the spin torque and the damping, 
which as a result stabilizes a steady precession. 
The validity of the developed theory was confirmed by performing the numerical simulation, 
showing a good agreement with the present theory. 


The authors would like to acknowledge 
T. Yorozu, H. Maehara, H. Tomita, T. Nozaki, K. Yakushiji, A. Fukushima, K. Ando, and S. Yuasa. 
This work was supported by JSPS KAKENHI Number 23226001. 




\end{document}